\documentclass{article}
\usepackage{spconf,amsmath,graphicx}
\usepackage{amsfonts}
\usepackage[percent]{overpic}
\usepackage{xcolor}

\usepackage{enumitem}
\setlist{nosep, leftmargin=14pt}

\usepackage{mwe} 

\title{CRC-SAM: SAM-Based Multi-Modal Segmentation and Quantification of Colorectal Cancer in CT, Colonoscopy, and Histology Images}
\name{
    Daniel Z. Lao \quad
    }
\address{
    Independent researcher \quad
}
\begin{document}
%
\maketitle
\begin{abstract}
We present CRC-SAM, a unified framework for colorectal cancer segmentation across colonoscopy, CT, and histopathology images. Unlike prior single-modality methods, CRC-SAM provides consistent, modality-agnostic segmentation throughout the clinical workflow. Built on MedSAM, it incorporates low-rank adaptation (LoRA) layers into a frozen encoder, enabling efficient domain transfer to underrepresented modalities with minimal trainable parameters. Experiments on MSD-Colon, CVC-ClinicDB, and EBHI-Seg demonstrate superior performance across modalities, outperforming state-of-the-art baselines and highlighting the effectiveness of lightweight LoRA adaptation for foundation-model-based colorectal cancer analysis.
\end{abstract}
\section{Introduction}
\label{sec:intro}

Colorectal cancer (CRC), the world’s third most common cancer (~10$\%$ of cases), arises from abnormal colon cell growth and often develops from benign adenomatous polyps. Because early stages are frequently asymptomatic, early detection and accurate diagnosis are critical for improving outcomes.

In the U.S., colorectal cancer screening typically begins with colonoscopy around age 50. When abnormalities are detected, CT imaging is used for tumor staging and metastatic assessment, followed by surgical resection and pathological analysis. Depending on disease stage, chemotherapy or radiation may be administered. Colonoscopy (CS), CT, and histopathology (PATH) together form the core modalities for colorectal cancer diagnosis and evaluation.

Accurate detection and segmentation of cancerous tissue are critical for quantitative colon cancer assessment. Automated methods improve diagnostic efficiency and consistency by reducing human subjectivity, with recent advances in deep learning enabling robust and reliable automatic analysis.

We present a unified framework for colorectal cancer segmentation that accurately delineates lesions across CS, CT, and PATH images. Due to substantial visual differences among modalities, specialist interpretations often vary, leading to inconsistent assessments. Building on the Segment Anything in Medical Images (MedSAM) framework \cite{ma2024segment,kirillov2023segment}, our approach leverages strong generalization capabilities to enable consistent, modality-agnostic segmentation and quantitative analysis across diverse imaging types. This streamlines diagnostic workflows and reduces inter-observer variability.
Our contributions can be summarized as:
\begin{itemize}[leftmargin=10pt]
\item We introduce CRC-SAM, a unified framework for colorectal cancer segmentation that achieves consistent and accurate lesion delineation across CS, CT, and PATH modalities, enabling reliable cross-modal clinical analysis.
\item We integrate a LoRA-based adaptation \cite{hu2022lora} into the frozen MedSAM encoder \cite{ma2024segment,kirillov2023segment}, enabling efficient domain transfer from CT/MRI-dominant representations to underrepresented imaging modalities. This design introduces only a small number of trainable parameters, substantially reducing computational cost while preserving strong cross-modal generalization.
\item We conduct extensive experiments on three public datasets, showing that CRC-SAM achieves SOTA performance across all modalities, demonstrating the effectiveness of lightweight adaptation for multi-modal medical image segmentation.
\end{itemize}

\section{Related work}
\label{sec:related_work}

Initial research on colorectal tissue segmentation was dominated by encoder–decoder CNNs such as U-Net \cite{ronneberger2015u}. UNet++ \cite{zhou2019unet++} advanced this paradigm by introducing dense and attention connections, and later studies further improved segmentation accuracy through enhanced boundary modeling and multi-scale feature representations.

Transformers have recently gained traction in medical image segmentation for their ability to capture global context. Hybrid CNN–Transformer models such as CTNet \cite{xiao2024ctnet} enhance multi-scale feature representation and attention integration for improved polyp segmentation. Despite these advances, challenges remain due to polyps’ subtle appearance and limited model generalization.

\begin{figure}[t]
    \vspace{-6pt}
    \centering
    \includegraphics[width=0.99\columnwidth]{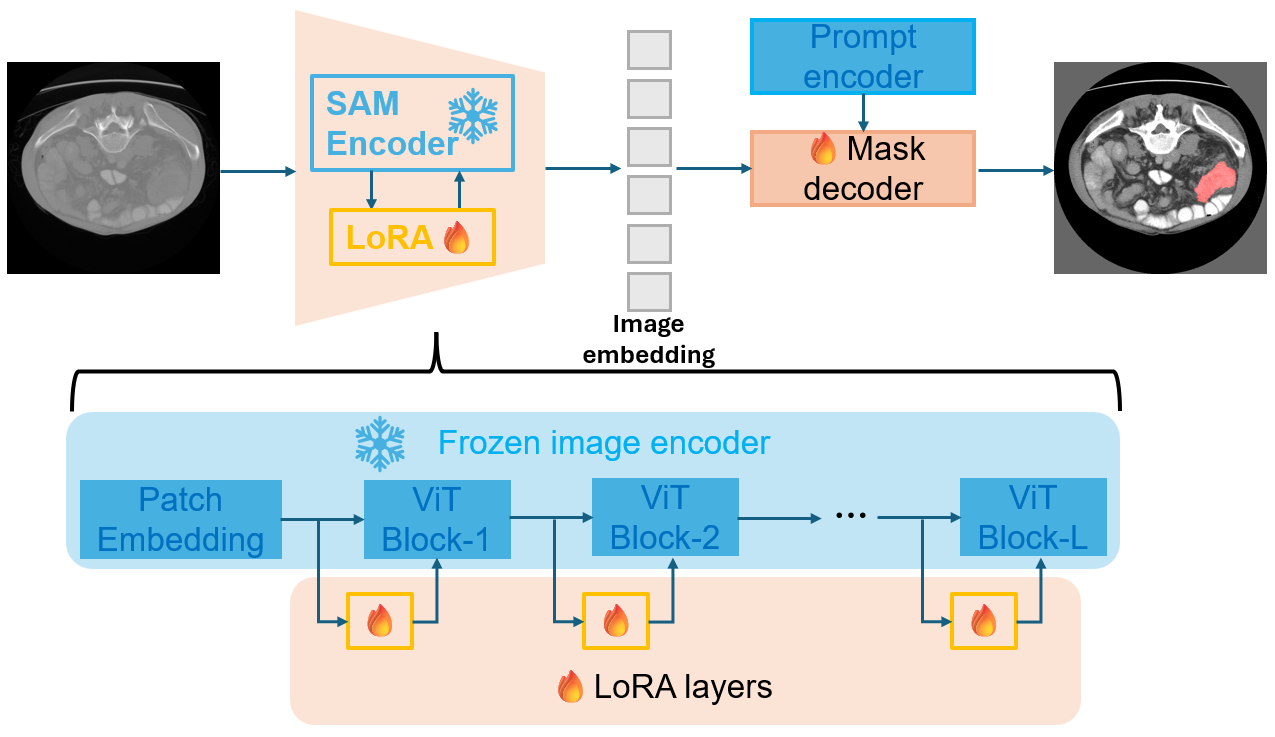}
    \vspace{-6pt}
    \caption{\small Overview of CRC-SAM. Built on MedSAM, the pipeline freezes the image and prompt encoders, integrates LoRA layers for medical feature adaptation, and fine-tunes the mask decoder for accurate colorectal cancer segmentation.}
    \vspace{-8pt}
    \label{fig:pipeline}
\end{figure}


\section{Method}
\label{sec:method}

\subsection{Overview}


Given an input image $x \in \mathbb{R}^{H \times W \times 3}$, the image encoder $f_{\theta}$ outputs embeddings $z = f_{\theta}(x) \in \mathbb{R}^{N \times D}$. The prompt encoder $g_{\phi}$ maps prompts $p$ to embeddings $e = g_{\phi}(p)$, which are fused by the mask decoder $h_{\psi}$ to produce the predicted mask $\hat{y} = h_{\psi}(z, e) \in [0,1]^{H \times W}$.


As shown in Figure~\ref{fig:pipeline}, CRC-SAM adopts the MedSAM architecture \cite{ma2024segment,kirillov2023segment}. The image encoder is frozen, while trainable LoRA-style bypass paths \cite{hu2022lora} are inserted into each Transformer block to project features into a low-rank space and reproject them back to the original feature dimensions. CRC-SAM supports fully automatic segmentation without user-provided prompts at inference, enabling streamlined clinical workflows. In the absence of prompts, MedSAM relies on a default embedding, which is fine-tuned in CRC-SAM to improve adaptation. The mask decoder is jointly fine-tuned to achieve accurate colorectal cancer segmentation across diverse imaging modalities.

\subsection{LoRA Integration in the Image Encoder}

Inside image encoder (Vision Transformer), each self-attention block has project $Q = X W Q, \quad K = X W K, \quad V = X W V$, where $X \in \mathbb{R}^{B \times N \times D}$ are the token embeddings (batch $\times$ patches $\times$ hidden dimension), and $W_Q, \, W_K, \,, W_V \in \mathbb{R}^{D \times D}$ are linear projection matrices. The attention output is computed as $A = \operatorname{softmax}\!\left(\frac{Q K^{\top}}{\sqrt{D}}\right) V$.

LoRA \cite{hu2022lora} adds a trainable low-rank update $\Delta W$ to frozen weights $W$: $W' = W + \Delta W = W + B A$, where $A \in \mathbb{R}^{r \times D}, B \in \mathbb{R}^{D \times r}, r \ll D \ \text{(e.g., } r = 4\text{)}$. Only $A$ and $B$ are trained; $W$ remains frozen.

Thus, the LoRA-enhanced attention projections become: $Q' = X (W_Q + B_Q A_Q), K' = X (W_K + B_K A_K), V' = X (W_V + B_V A_V)$. The low-rank term $B_Q A_Q$ injects task-specific adaptation (e.g., for colon texture) while keeping $W_Q$ fixed.

For a single attention block with LoRA \cite{hu2022lora}:
\begin{equation}
Z' = \operatorname{softmax}\!\left(\frac{X_Q X_K^{\top}}{\sqrt{D}}\right)X_V
\end{equation}
where $X_Q=X(W_Q + B_Q A_Q), X_K=X(W_K + B_K A_K), X_V=X(W_V + B_V A_V)$.

The difference from original MedSAM is the presence of $BA$ terms, which adjust internal feature representations toward the new domain (medical images) without touching the full base parameters.

After several such transformer blocks, the image encoder outputs:
\begin{equation}
z_{LoRA} = f_{\theta, LoRA}(x)
\end{equation}

where $f_{\theta, LoRA}$ denotes the ViT encoder modified with LoRA projections.

\subsection{Integration into mask decoder}

The mask decoder receives LoRA-adapted embeddings:
\begin{equation}
\hat{y} = h_{\psi}(z_{\text{LoRA}}, e)    
\end{equation}

If explicitly written:
\begin{equation}
\hat{y} = h_{\psi}\big(f_{\theta}(x) + f_{\text{LoRA}}(x; A, B),\, e\big)    
\end{equation}
where:
$f_{\theta}(x)$ denotes frozen encoder features (general visual knowledge from MedSAM\cite{ma2024segment,kirillov2023segment}),
$f_{\text{LoRA}}(x; A, B)$ denotes learned medical-specific feature adjustment.

Thus, LoRA \cite{hu2022lora} acts as a learned residual correction to the frozen encoder representation.

\subsection{Training objective}

The total loss $\mathcal{L}$ is applied only on the LoRA \cite{hu2022lora} and decoder parameters:
\begin{equation}
\mathcal{L} = \lambda_1\, \mathcal{L}_{\text{Dice}}(\hat{y}, y)
             + \lambda_2\, \mathcal{L}_{\text{CE}}(\hat{y}, y)    
\end{equation}
where $CE$ and $Dice$ denote cross entropy and Dice loss, respectively, with gradient flow:
\[
\nabla_{A,B,\psi}\, \mathcal{L} \neq 0, 
\qquad 
\nabla_{\theta,\phi}\, \mathcal{L} = 0.
\]

That is, only the LoRA matrices $(A, B)$ and the decoder $h_{\psi}$ are updated.

\section{experiments}
\label{sec:experiments}

\subsection{Datasets, Evaluation Metrics, and Model Parameters}
The proposed model is evaluated on CS, CT, and PATH modalities using the public datasets MSD-Colon, CVC-ClinicDB, and EBHI-Seg (Table~\ref{tab:dataset}).

MSD-Colon includes 190 annotated CT scans of colon cancer patients. In the absence of official test labels, the 126 training cases were split into 99 training and 27 testing samples.

CVC-ClinicDB includes 612 colonoscopy images (384 × 288), divided 8:2 into 490 training and 122 testing images.

EBHI-Seg contains 5,170 histopathology images spanning six tissue types. We use the adenocarcinoma subset (795 image–mask pairs), split 8:2 into 636 training and 159 testing samples.

Mean Dice Similarity Coefficient (mDSC) is adopted as the primary evaluation metric due to its widespread use and clinical relevance in medical image segmentation. Dice is closely related to IoU and precision/recall, offering a comprehensive measure of overlap-based performance. In contrast, boundary-based metrics such as HD95 are sensitive to annotation noise, which is common in CS and HIST datasets.

\begin{table}
    \centering
    \addtolength{\tabcolsep}{-5pt}
    \begin{tabular}{lcc}
        \hline
        \small{\textbf{Category}} & \small{\textbf{\#Parameters}} & \small{\textbf{\% of Total}}\\
        \hline
        \small{Total model parameters} & \small{93,735,472} & \small{100\%}\\
        \hline
        \textbf{\small{Total trainable}} & \textbf{\small{4,280,036}} & \textbf{\small{4.57\%}}\\
        --\small{LoRA (encoder)} & \small{221,184} & \small{0.24\%}\\
        --\small{Mask decoder} & \small{4,058,340} & \small{4.33\%}\\
        --\small{Default prompt embedding} & \small{512} & \small{$<0.1\%$}\\
        \hline
        \small{Total frozen} & \small{89,455,436} & \small{95.43\%}\\
        \hline
    \end{tabular}
    \caption{Model Parameter Breakdown (LoRA, r = 4)}
    \label{tab:model_parameter}
\end{table}

Table~\ref{tab:model_parameter} summarizes the model parameter breakdown and corresponding percentages, showing that only $4.57\%$ of the total parameters are trainable.

\begin{table}
    \centering
    \vspace{-6pt}
    \addtolength{\tabcolsep}{-5pt}
    \begin{tabular}{ccccc}
        \hline
        \small{Dataset} &  \small{Modality} & \small{Total} & \small{Training} & \small{Testing}\\ \hline
        \small{MSD-Colon} & \small{CT} & \small{126} & \small{99} & \small{27} \\
        \small{CVC-ClinicDB} & \small{CS} & \small{612} & \small{490} & \small{122} \\
        \small{EBHI-Seg} & \small{HIST} & \small{795} & \small{636} & \small{159} \\
        \hline
    \end{tabular}
    \caption{Datasets used in experiments}
    \label{tab:dataset}
\end{table}

\subsection{Results Qualitative Evaluation}

\begin{figure}[t]
  \centering
  \scalebox{0.8}{%
  \begin{minipage}[t]{\columnwidth}
    \centering
    \begin{overpic}[width=0.32\columnwidth]{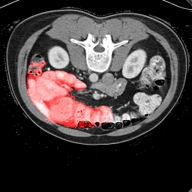}
      \put(2,90){\bfseries\color{white}DSC: 0.21}
    \end{overpic}%
    \hfill
    \begin{overpic}[width=0.32\columnwidth]{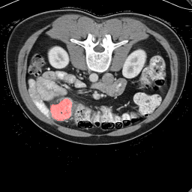}
      \put(15,90){\bfseries\color{white}DSC: 0.97}
    \end{overpic}
    \hfill
    \begin{overpic}[width=0.32\columnwidth]{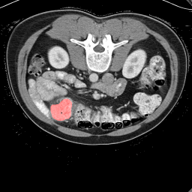}
    \end{overpic}
  \end{minipage}}
  \makebox[0.33\columnwidth][c]{\scriptsize MedSAM}%
  \hfill
  \makebox[0.33\columnwidth][c]{\scriptsize CRC-SAM}
  \hfill
  \makebox[0.33\columnwidth][c]{\scriptsize Ground Truth}
  \caption{\small Comparison of segmentation results: MedSAM (left), CRC-SAM (middle), and ground truth (right). Red regions indicate segmentation masks.}
  \label{fig:ct_medsam_compare}
\end{figure}




It is widely acknowledged that zero-shot MedSAM \cite{ma2024segment,kirillov2023segment} underperforms on downstream medical image segmentation tasks, particularly for underrepresented modalities. This limitation primarily stems from its training data, which is dominated by CT and MRI images whose visual characteristics, texture distributions, and intensity contrasts differ substantially from those of other medical imaging modalities. As illustrated in Figure~\ref{fig:ct_medsam_compare}, MedSAM frequently fails to accurately delineate colorectal cancer regions, highlighting its limited generalization beyond the training domain.

In contrast, CRC-SAM mitigates this limitation by incorporating an efficient LoRA-based adaptation mechanism \cite{hu2022lora} into the frozen MedSAM encoder. Although full fine-tuning of MedSAM can enhance performance, it requires updating a large number of parameters and incurs substantial computational cost. Our method achieves effective domain transfer by optimizing only a small set of low-rank parameters, enabling robust segmentation across diverse modalities with minimal computational overhead.

\begin{figure}[t]
  \centering
  \setlength{\tabcolsep}{1pt}
  \renewcommand{\arraystretch}{0.9}
  \scalebox{0.65}{%
  \begin{tabular}{cc}
    \begin{overpic}[width=0.46\columnwidth]{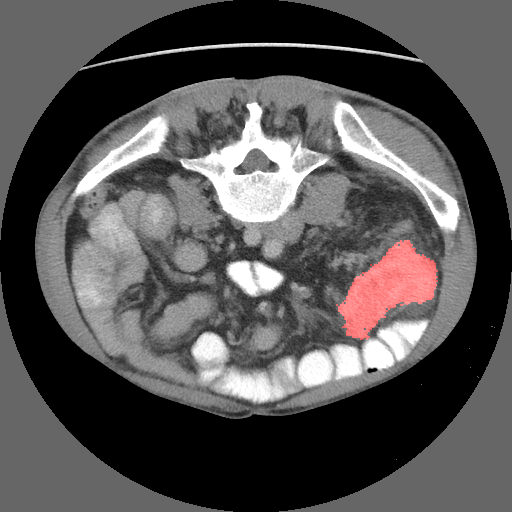}
      \put(2,90){\bfseries\color{white}\scriptsize DSC: 0.96}
    \end{overpic} &
    \begin{overpic}[width=0.46\columnwidth]{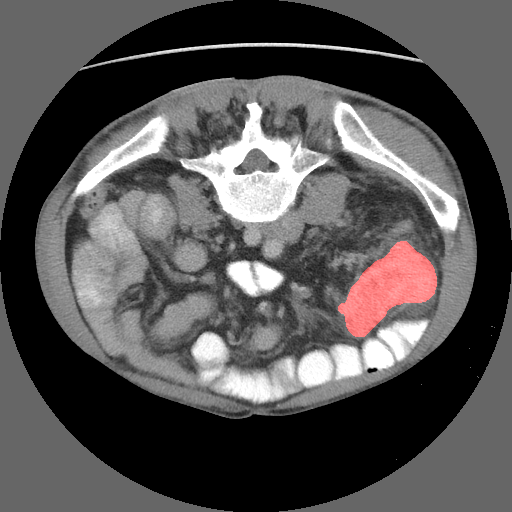}
    \end{overpic} \\
    \begin{overpic}[width=0.46\columnwidth]{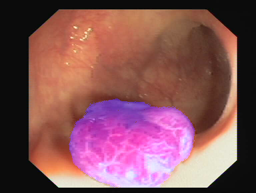}
      \put(2,65){\bfseries\color{white}\scriptsize DSC: 0.98}
    \end{overpic} &
    \begin{overpic}[width=0.46\columnwidth]{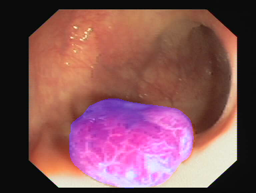}
    \end{overpic} \\
    \begin{overpic}[width=0.46\columnwidth]{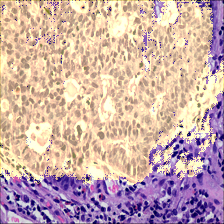}
      \put(2,90){\bfseries\color{black}\scriptsize DSC: 0.92}
    \end{overpic} &
    \begin{overpic}[width=0.46\columnwidth]{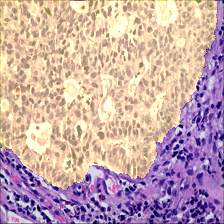}
    \end{overpic}
  \end{tabular}}
  \vspace{-4pt}
  \caption{\small Colorectal tumor segmentation examples. Predicted masks (left) and ground truth (right).}
  \label{fig:ct_pairs_singlecol}
  \vspace{-6pt}
\end{figure}

Figure~\ref{fig:ct_pairs_singlecol} shows representative colorectal cancer segmentation results produced by CRC-SAM, with predicted masks overlaid on the original images across three modalities, i.e., CT, CS, and HIST, along with the corresponding DSC scores.

\subsection{Quantitative Comparison with Prior Works}
\vspace{-3pt}

Our single model supports colorectal cancer segmentation across multiple medical imaging modalities. To the best of our knowledge, no existing method performs colorectal cancer segmentation jointly across all three modalities. As current SOTA approaches are modality-specific, we evaluate our framework separately against leading methods for each modality.

We evaluate CRC-SAM on CT scans from the MSD-Colon dataset, processing 3D volumes slice-by-slice and reconstructing volumetric predictions for evaluation. CRC-SAM is compared with UNet \cite{ronneberger2015u}, UNETR \cite{hatamizadeh2022unetr}, UNETR++ \cite{shaker2024unetr++}, MedNeXt, and SaSaMIM. As shown in Table~\ref{tab:quant_result_ct}, CRC-SAM achieves the highest mDSC among all methods.

Table~\ref{tab:quant_result_colonoscopy} compares CRC-SAM with four colorectal cancer segmentation methods on colonoscopy images. CRC-SAM consistently achieves the best performance, outperforming all competing approaches.

Finally, Table~\ref{tab:quant_result_histo} compares CRC-SAM with five methods for histopathology-based colorectal cancer segmentation, showing that our approach consistently achieves superior segmentation accuracy over all baselines.

Due to page limitations, extensive experimental results, including comparisons with prompt-based MedSAM, ablations on different LoRA ranks, unified versus modality-specific training, and runtime performance analysis, are omitted and will be made available on the project website.

\begin{table}[t]
    \centering
    \addtolength{\tabcolsep}{-2pt}
    \begin{tabular}{lc}
        \hline
        Method & mDSC (\%)$\uparrow$ \\
        \hline
        UNet \cite{ronneberger2015u}& 57.99 \\
        UNETR \cite{hatamizadeh2022unetr}& 52.19 \\
        UNETR++ \cite{shaker2024unetr++}& 56.80 \\
        MedNeXt & 57.61 \\
        SaSaMIM & \underline{64.73} \\
        CRC-SAM (Ours) & \textbf{90.55} \\
        \hline
    \end{tabular}
    \caption{Quantitative segmentation results on CT (MSD-Colon dataset). The best result is shown in \textbf{bold}, and the second-best is \underline{underlined}.}
    \label{tab:quant_result_ct}
\end{table}

\begin{table}[t]
    \centering
    \addtolength{\tabcolsep}{-2pt}
    \begin{tabular}{lc}
        \hline
        Method & mDSC (\%)$\uparrow$ \\ 
        \hline
        CFA-Net & 93.3 \\
        BA-Net & \underline{94.9} \\
        PPNet & 92.1 \\
        MMS-Net & 92.3 \\
        CRC-SAM (Ours) & \textbf{95.3} \\
        \hline
    \end{tabular}
    \caption{Quantitative segmentation results on colonoscopy (CVC-ClinicDB dataset). The best result is shown in \textbf{bold}, and the second-best is \underline{underlined}.}
    \label{tab:quant_result_colonoscopy}
\end{table}



\begin{table}[!t]
\centering
\vspace{-5pt}
\renewcommand{\arraystretch}{0.9}
\setlength{\tabcolsep}{3pt}
    \begin{tabular}{lc}
        \hline
        Method & DSC (\%)$\uparrow$ \\
        \hline
        UNet \cite{ronneberger2015u} & 78.66 \\
        UNet++ \cite{zhou2019unet++} & 79.57 \\
        ResAttUNet++ & 80.44 \\
        DRUNet & 81.70 \\
        RPAUNet++ & \underline{83.35} \\
        CRC-SAM (Ours) & \textbf{83.90} \\
        \hline
    \end{tabular}
    \caption{Quantitative segmentation results on histopathology (EBHI-Seg dataset). The best result is shown in \textbf{bold}, and the second-best is \underline{underlined}.}
    \label{tab:quant_result_histo}
\end{table}

\section{conclusion and discussion}
\label{sec:conclusion}

We propose a unified model for colorectal cancer segmentation across multiple medical imaging modalities, enabling consistent and objective analysis throughout the clinical workflow. Built on MedSAM \cite{ma2024segment,kirillov2023segment} with a lightweight LoRA-based adaptation \cite{hu2022lora}, the framework achieves accurate and efficient segmentation across CT, colonoscopy, and histopathology. Future work will explore larger and more diverse datasets to further evaluate generalization and cross-modality relationships.


\bibliographystyle{IEEEbib}
\bibliography{strings,refs}

\end{document}